\documentclass[preprint,12pt]{aastex}
\usepackage{apjfonts}
\usepackage{epsf}
\usepackage{graphicx}
\usepackage{amsmath}
\shortauthors{INOUE and SILK}
\shorttitle{Local Voids as the Origin of Large-angle CMB Anomalies}
\newcommand{\del}{\partial}

\newcommand{\K}{{\mathbf{k}}}
\newcommand{\x}{{\mathbf{x}}}

\newcommand{\f}{\frac}

\newcommand{\bb}{\bibitem}
\newcommand{\BF}{\begin{figure}\begin{center}}
\newcommand{\EF}{\end{center}\end{figure}}
\newcommand{\BE}{\begin{equation}}
\newcommand{\EE}{\end{equation}}
\newcommand{\BEA}{\begin{eqnarray}}
\newcommand{\EEA}{\end{eqnarray}}
\newcommand{\ti}{\textit}

\begin{document}
\title{Local Voids as the Origin of Large-angle Cosmic Microwave
Background Anomalies:  The Effect of a Cosmological Constant}
\author{Kaiki Taro Inoue and Joseph Silk}
\altaffiltext{1}{Department of Science and Engineering, 
Kinki University, Higashi-Osaka, 577-8502, Japan}
\altaffiltext{2}{University of Oxford, Department of Physics, 
Oxford, OX1 3RH, United Kingdom }


\begin{abstract}
We explore the large angular scale temperature anisotropies 
in the cosmic microwave background (CMB) 
due to homogeneous local dust-filled 
voids in a  flat Friedmann-Robertson-Walker 
universe with a cosmological constant. 
In comparison with the equivalent dust-filled void model 
in the Einstein-de Sitter background, we find that the 
anisotropy for compensated asymptotically expanding 
local voids can be larger because second-order effects
enhance the linear integrated Sachs-Wolfe (ISW) effect. 
However, for local voids that expand sufficiently faster than the 
asymptotic velocity of the wall, 
the second-order effect can suppress the fluctuation due to the 
linear ISW effect. 
A pair of quasi-linear compensated asymptotic local voids with radius 
$(2-3)\times 10^2 ~h^{-1}$ Mpc and a matter density contrast $\delta_m 
\sim -0.3$ can be observed as cold spots with a temperature anisotropy
$\Delta T/T\sim O(10^{-5})$ that might help explain the observed large-angle
CMB anomalies. We predict that the associated  anisotropy in the local 
Hubble constant in the direction of  the 
voids could be as large as a few percent.
   
\end{abstract}


\section{Introductio}
Recently, there has been mounting evidence that 
the statistical isotropy in the 
large-angle cosmic microwave background (CMB) anisotropy 
may be broken (Tegmark et al. 2003;
Copi et al. 2004;  Chiang et al. 2004; 
de Oliveira-Costa et al. 2004; Eriksen et al. 2004; 
Hansen et al. 2004; Larson \& Wandelt 2004; Park 2004; 
Schwarz et al. 2004; Vielva et al. 2004; Cruz et al. 2005, 2006 ). 
Although the significance of any one of 
the findings is at most 3$\sigma$, the accumulation of anomalies
hints at  the possibility of  the need for new physics to be added to the standard scenario
(e.g., Luminet et al. 2003; Jaffe et al. 2005).

To explain the origin of the anomalies, it has been suggested that 
the large-angle CMB anisotropy 
is affected by  local inhomogeneities (Moffat 2005; 
Tomita 2005a,2005b; Vale 2005; Cooray \& Seto 2005; Raki\'c et al. 2006). 
However, none of these explanations has succeeded in explaining the 
specific features 
of the anomalies, namely, the octopole planarity, the alignment
between the quadrupole $(l=2)$ and 
the octopole ($l=3$) (Tegmark et al. 2003), and the alignment 
between the low-$l$ multipoles $(l=2+3)$ with the equinox and the 
ecliptic plane (Copi et al. 2004). For instance, if one 
applies a model in which the Local Group is falling into 
the center of the Shapley supercluster, 
the discrepancy between the model prediction and the 
observed data becomes even worse (Raki\'c et al. 2006). 

Inoue \& Silk (2006)  explored the possibility 
that the CMB is affected by a small number of compensated local
dust-filled voids. In the Einstein-de Sitter (EdS) background, 
we found that a pair of voids
with  radius $~3\times 10^2~h^{-1}$ Mpc and density contrast 
$\delta=-0.3$ can account  for the planarity/alignment 
features in multipoles with $l=2$ and $l=3$. 
The cold spot in the Galactic southern 
hemisphere, anomalous at the 
$\sim 3\sigma$ level, can also be explained by such a large 
void at $z\sim 1$.    

In this paper, we study the signature of compensated local 
dust-filled voids on the CMB in a  flat 
Friedmann-Robertson-Walker (FRW) model with a cosmological
constant $\Lambda$. 
The signature of voids in the FRW universe has been 
extensively studied in the literature (Thompson \& Vishniac 1987; 
Mart\'inez-Gonz\'alez \& Sanz 1990; Panek 1992; 
Arnau et al. 1993; M\'esz\'aros 1996; Fullana et al. 1996; 
Vadas 1998; Griffiths et al. 2003). 
However, none of the above discussions have considered
the effect of the cosmological constant which accelerates the 
cosmic expansion at late times.  
In contrast to the void model in the 
EdS background, we expect an extra contribution 
from the linear integrated Sachs-Wolfe (ISW) effect due to 
potential decay at recent epochs $z<1$. It is of considerable
importance to study whether the second-order effect 
observed in the equivalent void model in the EdS background 
enhances the linear ISW effect or not (Tomita 2005a, 2005b). 
In a similar manner to our previous work (Inoue \& Silk 2006), we use
the thin-shell approximation for describing the wall.
In what follows, we assume that we are outside the voids. 
In section 2, we derive analytic formulae for the temperature anisotropy
due to homogeneous dust-filled voids
in the FRW model with a cosmological constant $\Lambda$, and we
study the signature of such voids on the CMB.
In section 3, we explore a model of local dust-filled voids that agrees 
with the observed anomalies on large angular scales.  
In section 4, we estimate the mean contribution from 
the local dust-filled voids in the cold dark matter (CDM) cosmology. 
In section 5, we summarize our results and discuss some unresolved 
issues.

\section{Dust-filled void model}

\subsection{Cosmic expansion}
We consider a flat Friedmann-Robertson-Walker
(FRW) model with a cosmological constant $\Lambda$
as the background universe. 
In what follows, we use units 
where the light velocity $c$ is normalized to 1. 
In
spherical coordinates $(r,\theta,\phi)$, the flat background FRW metric
can be written as 
\BE
ds^2=-dt^2+R^2(t)(dr^2+r^2 d \Omega^2),
\EE
where $t$ is the time, $R$ denotes the scale factor, and
$d \Omega^2$ is the metric for a unit sphere.
Next, we consider a 
homogeneous spherical dust-filled void 
with a density contrast $\delta_m<0$. 
We assume that the size of the void is 
sufficiently smaller than the Hubble radius $H^{-1}$. Then, the metric
of the homogeneous void 
centered at the origin can be approximately described by 
the hyperbolic FRW metric as 
\BE
ds^2=-dt'^2+R'^2(t')(dr'^2+ R_c^2 \sinh^2(r'/R_c)d \Omega^2),
\EE
where $t'$ is the time, $R'$ denotes the scale factor, 
and $R_c$ is the comoving curvature radius.
The scale factor $R$ 
for the flat FRW background at 
time $t_1$  can be expanded up to  third order in 
$\Delta t=t_2-t_1$ in terms of the 
deceleration parameter $q_2$ and the jerk parameter $j_2$ 
\BE
q_2\equiv -\f{\ddot{R}R}{\dot{R}^2}
\bigg |_{t_2},~~~~ j_2\equiv -\f{\dddot{R}R^2}{\dot{R}^3}
\bigg|_{t_2}, \label{eq:3}
\EE
as 
\BE
R(t_1) \approx  1-H_2 \Delta t -\f{q_2}{2}(H_2 \Delta t)^2
+\f{j_2}{6}(H_2 \Delta t)^3
\label{eq:4}
\EE, where a dot means the time derivative $d/dt$, $H_2\equiv H(t_2)$, 
and $R(t_2)\equiv 1$. 
For a FRW universe with matter and $\Lambda$, we have 
$q=(\Omega_m-2 \Omega_\Lambda)/2$ where $\Omega_m$ and $\Omega_\Lambda$
are the density parameters for the matter and the cosmological
 constant $\Lambda$, and the jerk is $j=-(\Omega_m+\Omega_\Lambda)$. 
The density parameter and the Hubble
parameter (denoted by primes) for the void are written in terms of the 
matter density $\rho_m$, the matter density 
contrast $\delta_m=\delta \rho_m/\rho_m$, and 
the Hubble parameter contrast $\delta_H$ as 
\BE
\Omega'_{m}=\f{\rho_{m}(1+\delta_m)}{H^2 (1+\delta_H)},
~~~H'=(1+\delta_H)H.
\label{eq:5}
\EE
The deceleration and jerk parameters for the void are
\BEA
q'&=&\f{1}{2(1+\delta_H)^2}\biggl[\Omega_m(1+\delta_m)-2 \Omega_\Lambda
  \biggr],
\nonumber
\\
j'&=&\f{1}{(1+\delta_H)^2}\biggl[\Omega_m(1+\delta_m)+ \Omega_\Lambda
  \biggr].
\label{eq:6}
\EEA
Equations (\ref{eq:3},\ref{eq:4},\ref{eq:5},\ref{eq:6}) give the scale factor
$R'(t')$ for the hyperbolic FRW void. 
From the Friedmann equation, the cosmological time for
the flat universe that consists of 
matter and the cosmological constant $\Lambda$ is
\BE
t=s H^{-1}, ~~~s=\f{2}{3\sqrt{1-\Omega_m}} \ln 
\biggl[\f{\sqrt{1-\Omega_m}+1}{\sqrt{\Omega_m}}   \biggr]. 
\label{eq:7}
\EE

\subsection{ Time solution}

In what follows, we calculate the evolution of the internal 
time $t'$ for an  expanding 
void with a peculiar velocity 
in terms of the external time $t$ up to 
order $O[(r_v/H^{-1})^3]$, where the comoving radius of the 
void is expressed as $r_v(t)$ in the external coordinates
and as $r'_v(t')$ in the internal coordinates.
To connect the two metrics at the shell,  we require
the boundary conditions:
\BE
(R(t)r_v(t))^2=(R'(t')r'_v(t'))^2\biggl(1+\f{R'(t')^2 r'_v(t')^2}{3 R_c^2}\biggr), 
\label{con1}
\EE
and
\BE
-dt^2+R^2(t)dr^2=-dt'^2+R'^2(t')dr'^2. \label{con2}
\EE
where we have assumed that $r'\ll R_c$.
Up to order $O[(r_v/H^{-1})^3]$, the curvature term in 
equation (\ref{con1}) is negligible. Then equation 
(\ref{con1}) and  (\ref{con2}) yield
\BE
dt'=dt\biggl
(1-R\dot{R}r_v\dot{r_v} \delta_H+\f{1}{2}\dot{R}^2 r_v^2 \delta_H^2
\biggr). \label{dt'1}
\EE
Let us assume that the expansion of the thin shell (=wall)
in the external coordinates is expressed as $r_v(t)\propto t^\beta$, where $\beta$
is a constant. Equation (\ref{dt'1}) can be explicitly
written as
\BE
dt'=
\biggl(1- \f{\beta \delta_{H}}{s} \xi^2+
\f{1}{2}\delta_{H}^2 \xi^2 \biggr)dt, \label{eq:11}
\EE
where we define 
\BE
\xi \equiv \f{r_v(t)R(t)}{H^{-1}}. \label{eq:12}
\EE
Note that $\delta_H$ and $\xi$ are functions 
of $t$ rather than $t'$. As we shall see in the 
subsequent analysis, the anisotropy turns out 
to have a leading order of $\xi^3$. Therefore, 
omitting the curvature term in equation ($\ref{con1}$)
\textit{in the process of deriving} $dt'$ can be justified\footnote{The
curvature correction yields additional terms of
order $O(\xi^4)$ in equation (\ref{eq:11}). }.
Using equations
(\ref{eq:3},\ref{eq:4},\ref{eq:5},\ref{eq:6},\ref{eq:7}), and
(\ref{eq:12}), equation (\ref{eq:11}) can be integrated to yield 
the finite time difference $\Delta t'\equiv t'_2-t'_1$,
\BE
\Delta t'=c_1 \Delta t+c_2 (\Delta t)^2+c_3 (\Delta t)^3, 
\EE
where $c_i$ for $i=1,2,3$ are functions of $\beta$ and $\delta_H$
at $t=t_2$. 

\subsection{Temperature anisotropy}
We denote quantities at the time  the photon enters
the void and those at the time  the photon leaves
by the subscripts ``1'' and ``2'', respectively.
Primes denote quantities measured by a comoving observer
in the interior coordinate system, and the unprimed quantities
are measured by a comoving observer in the background universe
just out of the shell of the void (see figure 1 in Inoue \& Silk 2006). 

To calculate the energy loss, we apply two local
Lorentz transformations at each void boundary. 
The first is to convert the photon four-vector momentum 
in the comoving frame in the background universe
to the frame in which the shell is at rest. 
The second is to convert it to the frame in the comoving frame
inside the void.

The  four-vector momentum of the photon that enters the void is
\BE
\bold{k}_1 \equiv E_1\left(
\begin{array}{c}
1 \\ \cos\psi_1 \\ \sin\psi_1 \\ 0
\end{array}\right),
\EE
where $\psi_1$ is the angle between the normal vector of the void shell 
and the spatial three-vector of the momentum of the photon
that enters the void. 
After the photon passed the shell,
the four-vector is converted to 
\BEA
\bold{k}'_1 
&\equiv&
 E_1'\left(
\begin{array}{c}
1 \\ \cos\psi'_1 \\ \sin\psi'_1 \\ 0
\end{array}
\right)
\\
&=& E_1\left(
\begin{array}{c}
\gamma_1\gamma_1'[1+(v_1-v_1')\cos\psi_1-v_1v_1'] \\
\gamma_1\gamma_1'[\cos\psi_1+(v_1-v_1')-v_1v_1'\cos\psi_1] \\
\sin\psi_1 \\ 0
\end{array}\right), \label{k1'}
\EEA
where $v_1$ and $v_1'$ are the velocities of the void shell
at the time $t=t_1$ and $\gamma$-factors are defined as 
$\gamma_1=1/(1-v_1^2)^{1/2}$ and $\gamma_1'=1/(1-v_1'^2)^{1/2}$.  
When the photon reaches the far edge of the shell, the 
four-vector momentum becomes 
\BE
\bold{k}_2' \equiv \f{R_1'}{R_2'}E_1' \left(
\begin{array}{c}
1 \\ \cos\psi_2' \\ \sin\psi_2' \\ 0
\end{array}\right),
\EE  
where $\psi_2'$ is the angle between 
the normal vector of the void shell 
and the spatial three-vector of the momentum of the photon
that leaves the void. 

As the photon leaves the shell, the four-vector momentum
is converted to 
\BEA
\bold{k}_2 
&\equiv&
 E_2\left(
\begin{array}{c}
1 \\ \cos\psi_2 \\ \sin\psi_2 \\ 0
\end{array}
\right)
\\
&=&
\f{R_1'}{R_2'}E_1'\left(
\begin{array}{c}
\gamma_2\gamma_2'[1+(v_2-v_2')\cos\psi_2'-v_2v_2'] \\
\gamma_2\gamma_2'[\cos\psi_2'+(v_2-v_2')-v_2v_2'\cos\psi_2'] \\
\sin\psi_2' \\ 0
\end{array}\right).\label{k2}
\EEA
The velocities of the void are   
\BEA
v_i&=&R \f{dr_v}{dt}\bigg |_{t=t_i}=\beta R \f{r_v}{t_i} 
\label{velocity-void}
\\
v_i'&=& R'\f{dr'_v}{dt'}\bigg |_{t'=t_i'},
\EEA
where $i=1,2$. From the connection conditions
(\ref{con1}), (\ref{con2}), and equation (\ref{dt'1}), terms up to 
order $\cal{O}(\xi^3)$, $v_i'$ can be calculated as
\BEA
v_i'(t_i)&=&\biggl(1-\f{\kappa_i}{2}\xi_i^2 \biggr)
\biggl\{\biggl ( 
1+ \f{\beta \delta_{Hi}}{s_i} \xi_i^2-
\f{1}{2}\delta_{Hi}^2 \xi_i^2
\biggr)\bigl(v_i+\xi_i \bigr)
\nonumber
\\
&-&\xi_i (1+\delta_{Hi})\biggr\}, 
\label{vi'}
\EEA
where $\xi_i\equiv \xi(t_i)$and
$\delta_{Hi}\equiv \delta_H(t_i), \kappa_i\equiv (1+\delta_{Hi})^2
-(1+(1-\Omega_{\Lambda}(t_i))\delta(t_i))$.
Using the Friedmann equation, 
the parameters at $t=t_1$ can be written in terms of those
at $t=t_2$ as
\BE
\f{1+\delta_1}{1+\delta_2}=\f{H_2^2}{H_1^2}
\biggl(\f{R_2'}{R_1'}\biggr)^3 \biggl(
\f{1-\Omega_{\Lambda}(t_2)}{1-\Omega_{\Lambda}(t_1)(H_2/H_1)^2}  
\biggr),
\EE
and
\BE
\f{\kappa_1}{\kappa_2}=\f{H_2^2}{H_1^2}
\biggl(\f{R_2'}{R_1'}\biggr)^2.
\EE
From the geometry of the void in the internal 
comoving frame, at the order $O(\xi^3)$, 
the relation between the void radius 
$r_{v1}'$ and $r_{v2}'$ is approximately given by
\BE
r_{v1}' \sin{\psi_1'}\approx r_{v2}' \sin{\psi_2'}. 
\label{geom1}
\EE
The relation between the time $t_1$ and $t_2$
can be written as
\BE
\int_{t_1'}^{t_2'} \f{dt'}{R'(t')} \approx r_{v1}' \cos \psi_1'+r_{v2}' \cos
\psi_2'. \label{geom2}
\EE
The energy loss suffered between times $t_1$ and $t_2$ by a
CMB photon that does not traverse the void is
\BE
\biggl(\f{E_2}{E_1}\biggr)_{\textrm{no void}}=\f{1+z_2}{1+z_1},
\EE
where $z_1$ and $z_2$ are redshift parameters corresponding to
$t_1$ and $t_2$, respectively.
The ratio of the temperature change for photons that traverse the void 
to that for photons that do not traverse the void is
\BE
\f{\Delta T}{T}=\biggl( \f{E_2}{E_1}\biggr )_{\textrm{void}}
\f{1+z_1}{1+z_2}-1. \label{dToverT}
\EE
Equations (7-28) and can be solved recursively using $\xi_2$ as 
a small parameter.

\begin{figure*}
\epsscale{0.5}
\plotone{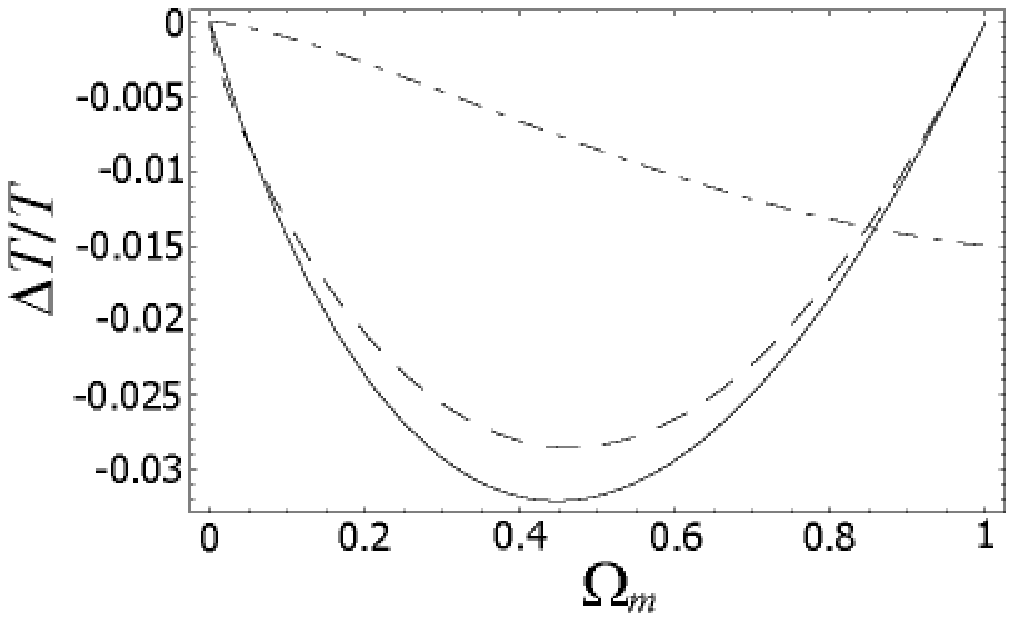}
\caption{Amplitude of temperature anisotropy for a compensating 
void with $\delta_m=-0.3$ as a function of the mass density $\Omega_m$
with the contribution from the term proportional to 
$\delta_m$ in equation (\ref{eq:29}) (solid curve), 
in equation (\ref{B6}) (dashed curve)  and that from the term
proportional to $\delta_m^2$ in equation (\ref{eq:29}) (dash-dotted curve).
We assume $\xi_2 \equiv r_{v2}R_2/H_2^{-1}=1$ and $\psi_2=0$.  }
\label{fig:f1}
\end{figure*}

After a lengthy calculation, neglecting the terms 
of order $O(\kappa^2)$\footnote{It turns out that the terms of 
order $O(\kappa^2)$ are actually 
absent in $\Delta T/T$ if one writes $\delta_H$ in terms of $\kappa$
 and $\delta_m$. The next leading order is $O(\kappa^3)$.}, we find
\BEA
\f{\Delta T}{T}&=&\f{1}{3 }\biggl[\xi_2^3 \cos{\psi_2}
\Bigl(-2 \delta_H^2-\delta_H^3+(3+4 \delta_m)\delta_H \Omega_m
\nonumber
\\
\nonumber
&+&\delta_m \Omega_m(-6 \beta/s+1)+
(2\delta_H^2+\delta_H^3+\delta_m \Omega_m
\\
&+&(3+ 2\delta_m)
\delta_H \Omega_m)\cos{2 \psi_2}\Bigr)\biggr]
, \label{result}
\EEA
where $\delta_H$, $\delta_m$, $\beta$, $s$, $\Omega_m$
are evaluated at $t=t_2$. 
For the matter-dominated EdS
Universe, $\delta_H=-\delta_m/3$, $\Omega_m=1$, and $\Lambda=0$,
we recover formula (34) from (Inoue \& Silk 2006).
Equation (\ref{result}) is the generalized formula
for a spherical homogeneous dust-filled void in the
flat-$\Lambda$ FRW background.
Plugging in $\beta=-\Omega_m^{0.6} \delta_m/6$
for a compensating void (Sakai 1995; Sakai et al. 1999)
\footnote{This approximation 
formula was originally derived for linear 
voids in low-density $(\Omega_0<1)$ FRW models.  However,
from numerical analyses, it turns out that it can also be 
be applied to voids in 
flat-$\Lambda$ FRW models if $\delta_m\ll 1$ (Sakai 1995). 
  }
and the Hubble parameter contrast in 
linear order, $\delta_H=\Omega_m \delta_m (1+f^{-1}(w))/2$
(see the derivation and the definition of $f(w)$ in the appendix),
into equation (\ref{result}), the anisotropy formula for compensated 
voids is 
\BEA
\f{\Delta T}{T}
&=&\f{1}{3}\delta_m \Omega_m \xi_2^3 \cos^3(\psi_2)
\biggl(2+3\Omega_m -\f{5 \Omega_m^{2/3}}{F}  \biggr)
\nonumber
\\
\nonumber
&+&
\f{\delta_m^2 \Omega_m^{1/3}\xi_2^3 \cos(\psi_2)}
{54 F^2}
\biggl[-50 \Omega_m \sin^2(\psi_2)
\\
\nonumber
&-&30 \Omega_m^{4/3} F(1+
2 \cos(2 \psi_2))+
9\bigl(2\Omega_m^{1.27}/s+3 \Omega_m^{5/3}
\\
&+&
3 \Omega_m^{5/3}
\cos(2 \psi_2)\bigr)F^2\biggr]+\xi_2^3 O(\delta_m^3), \label{eq:29}
\EEA
where
\BE
F\equiv _2F_1\biggl[5/6,1/3,11/6,-w \biggr], \label{eq:30}
\EE
and $w$ is the effective equation-of-state parameter
at time $t_2$. In contrast to the void model in the
EdS background, there is an additional
term which is linear in the matter density contrast $\delta_m$.
As shown in figure \ref{fig:f1}, it can be interpreted as the
contribution from the linear ISW effect,
which comes from the change in the gravitational 
potential as the photon crosses the void. 

For a void at redshift $z$ centered at an angular diameter distance 
$d$ from the Earth, the angle $\psi_2$ between the light ray 
and the normal vector on the void surface at the time
the photon leaves the shell is written
in terms of the size/distance parameter $b \equiv R(z) r_v/d$
and the angle $\theta$ between the light ray and the line of sight
to the void center as
\BE
\psi_2(b,\theta)\approx \sin^{-1}(\sin\theta/b). \label{psi}
\EE
Equations (\ref{eq:29}),(\ref{eq:30}), and (\ref{psi}) 
give the anisotropy for compensating 
voids at arbitrary redshift $z$. 
For a void at $z=0$ in the flat-$\Lambda$ FRW universe
with $\Omega_{m,0}=0.24$, (Spergel et al. 2006) we have $w_0=\Omega_{m,0}-1=-0.76$. 
Then equations (\ref{eq:29}) and (\ref{eq:30}) yield
\BEA
\f{\Delta T}{T}&\approx&\xi_0^3 \cos{\psi_2}
\biggl[0.087 (\cos^2{\psi_2}) \delta_{m,0}
\nonumber
\\
&-&\bigl(0.021+0.018 
\cos^2{\psi_2} \bigr)\delta_{m,0}^2\biggr],
\EEA
where $\xi_0$ and $\delta_{m,0}$ are evaluated at $t=t_0$.
Plugging $\psi_2=0$ $(\theta=0)$ in equation (\ref{eq:29}),
we have the anisotropy  
\BE
\f{\Delta T}{T} \approx 
(0.087\delta_{m,0}
-0.039\delta_{m,0}^2) \xi_0^3 \label{eq:DeltaT}
\EE
toward the center of the void. From equation (\ref{eq:DeltaT}),
we conclude that the contribution from the 
second-order effect, \footnote{Here we do not call this the 
``Rees-Sciama effect'' (Rees \& Sciama 1968) because we deal with 
the relativistic second-order effect in which the curvature effect
is no longer negligible. } which is proportional  to $\delta_m^2$ 
coherently enhances the contribution from the linear ISW effect
provided that the second-order correction for the void expansion 
and the wall expansion (which we discuss later) is negligible. 
In this case, we would observe a 
cold spot in the direction of the void.

To generate an anisotropy $\Delta T/T=10^{-5}$, the 
void radius should be
\BE
r_{v}(z=0)\approx 
2 \times 10^2  (|\delta_{m,0}|/0.2)^{-1/3} 
~~h^{-1}
\textrm {Mpc},
\EE
where $H^{-1}_0=3000~h^{-1}$Mpc. In comparison with 
the equivalent void model in the EdS background,
the void radius can be smaller because of the coherent 
contribution owing to the linear ISW effect and the
second-order effect.

\begin{figure*}
\epsscale{1.}
\plotone{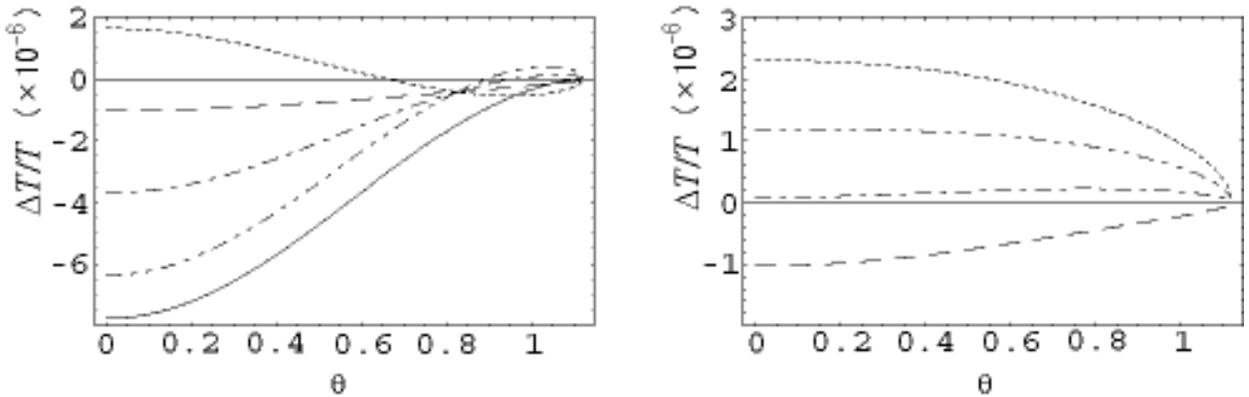}
\caption{Second-order effects.
Temperature anisotropy profiles owing to the second-order effect 
are plotted as a function of the subtending angle $\theta$. 
We assume a compensating void with a radius 
$r_v=200h^{-1}$ Mpc, $\delta_m=-0.3$, and the size/distance parameter 
$b=0.9$. {\ti{Left}}:($\epsilon,\eta$)=(-0.3,0) (dotted curve), 
 ($0,0$) (dashed curve), ($0.3,0$) (dash-dotted curve), and
($0.6,0$) (dash-dotted-dotted curve). 
{\ti{Right}}:($\epsilon,\eta$)=($0,0.9$) (dotted curve), 
 ($0,0.6$) (dash-dot-dotted curve), ($0,0.3$) (dash-dotted 
 curve), and ($0,0$) (dashed curve). The solid curve in the left panel 
denotes the temperature anisotropy profile owing to the linear ISW effect.}
\label{fig:f2}
\end{figure*}

In order to study the 
second-order corrections, we introduce
two nonlinear parameters ($\epsilon, \eta$) that control
the void and the wall expansion (see also Inoue \& Silk 2006).  
These are defined as
\BE
\delta_H=\f{1+f^{-1}(w)}{2}\Omega_m \delta_m-\epsilon \delta_m^2,
\EE
and
\BE
\beta=-\f{\Omega_m^{0.6}}{6} \delta_m+\eta \delta_m^2.
\EE
If the matter inside the void near the boundary 
expands with the asymptotic velocity of the wall, i.e., 
$\delta_H=\beta$, then the non-linear 
parameter $\epsilon$ can be written as 
\BE
\epsilon=\f{1}{6 \delta} \biggl(\Omega_m^{3/5}+ 
3\Omega_m -6 \delta \eta -5 F^{-1} \Omega_m^{2/3} 
\biggr).
\EE
Assuming $\eta=0$, we have $\epsilon>0$ for $\Omega_m>0$.
For low-density universes 
$\Omega_m =0.2-0.3$, we have $\epsilon=(0.07-0.09)|\delta|^{-1} $.
As shown in figure \ref{fig:f2},
an increase in $\epsilon$ leads to 
further redshifts of the photons due to
reduction in 
the expansion rate inside the void. 
Therefore, for voids that expand
with the asymptotic velocity, the 
second-order effect always enhances the linear ISW effect.
On the contrary, for $\epsilon<0$, the second-order 
effect leads to further blueshifts of the photons.  
Therefore, for voids that expand sufficiently faster than the 
asymptotic velocity of the wall (i.e. $\delta_H>\beta$), 
the second-order effect can reduce the redshift of photons
due to the linear ISW effect. 
Furthermore, an increase in the velocity of the wall (i.e. 
$\eta>0$) also leads to a suppression of the linear ISW effect 
because the photon is more Doppler blueshifted (see 
figure \ref{fig:f2}, {\it{right}}).  
Thus, the net redshift/blueshift of photons 
upon leaving the void depends on whether it is
asymptotically evolving or not.   
In what follows, we define asymptotic voids 
as those with $\delta_H=\beta$ and $\eta=0$. 
\section{Origin of anomalies} 
\begin{figure*}
\epsscale{0.7}
\plotone{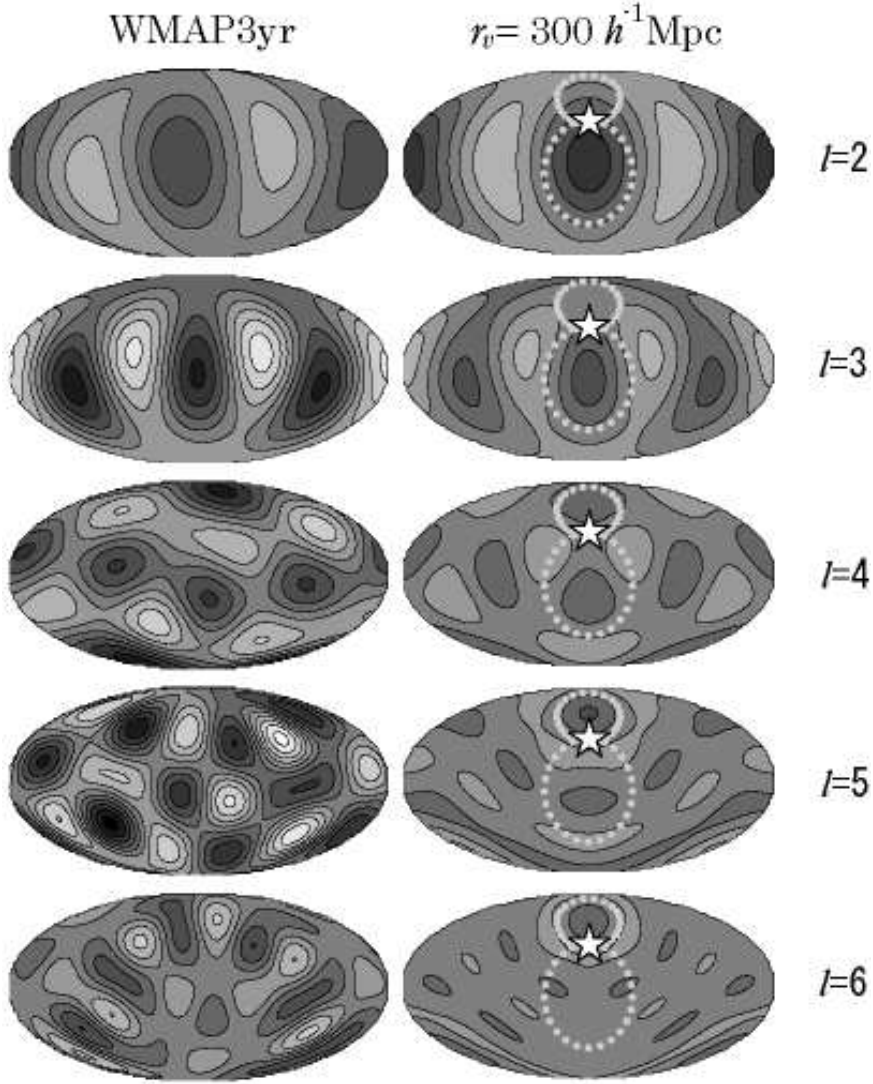}
\caption{Mollweide projection maps 
for the cleaned CMB from the WMAP third-year data (Park et al. 2006)
 ({\it{left}}) and those for the temperature anisotropy owing to a
pair of identical voids with radius $r_v=300 h^{-1}$ Mpc 
({{\it right}}) for modes $l=2-6$. We assume 
a flat-$\Lambda$ FRW universe with $\Omega_{m,0}=0.24$. 
The boundaries of the void hemispheres are denoted by small 
light-gray disks. Note that the identical voids are 
tangential to each other and the apparently larger void is nearer 
to us. We also assume that we are outside the pair of voids. 
The size/distance ratios of the 
two voids are assumed to be $b=0.91$ and $b=0.46$.  
The separation angle of the void centers 
is $\theta_s=63^{o}$ and the non-linear parameters are chosen to be
$(\epsilon,\eta)=(0.27,0)$, which correspond to an asymptotic void 
$(\delta_H=\beta$). The gray scale denotes the temperature fluctuations 
in which the maximum absolute value is set to $50\mu$K. 
The north pole is aligned to the $z-$axis for which the
angular dispersion of the quadrupole
plus the octopole around the $z-$axis (de-Oliveira Costa et al. 2004) 
is maximal in the direction
$(l,b)=(-110^\circ,60^\circ)$, which is close to the 
dipole direction at $(l,b)=(-96^\circ,48^\circ)$. 
The coordinate center is located 
at $(l,b)=(-30^\circ,-30^\circ)$.  The Shapley supercluster 
is located in the direction of the point where the 
two voids contact each other (star). } \label{fig:f3}
\end{figure*}

In order to explain the low-$l$ anomalies, 
we have proposed a model that consists of
a pair of identical local voids in the direction  
$(l,b)=(-30^\circ,-30^\circ)$ at $z \sim 0.05$ (Inoue \& Silk 2006). 
In the flat-$\Lambda$ background, a similar 
model can work as well. For illustrative purpose,
we have simulated skymaps for models that consist of identical
asymptotic voids with a density contrast $\sim -0.3$ 
that are tangent to each other. Note that 
we are assumed to be just outside 
one of the voids. As the background cosmology,
we have assumed a flat-$\Lambda$ FRW model with $\Omega_{m,0}=0.24$. 
As shown in figure \ref{fig:f3}, the planar feature $l=2,3,6$ 
and the alignment feature $l=2,3$ can be 
well reproduced in our model that consists of 
identical voids with radius $(2-3)\times10^2 h^{-1}$ Mpc.

Although we still need an additional contribution 
that does not completely blur the fluctuation
pattern by the voids, the chance 
of alignment between the quadrupole and octopole
becomes large in our void model since otherwise
no correlation between the two modes is expected.

In order to study the large-angle correlation 
feature, we also calculated the two-point  
correlation function in real space $C(\theta)$ (figure 4).
For $r_v=300 h^{-1}$ Mpc, the amplitude for the
separation angle $\theta>30^\circ$ is found to be 
$C(\theta)\sim 100 \mu \textrm{K}^2$ which is marginally
consistent with the observed WMAP 3-year 
values (see Spergel et al. 2006).

Recently, it has been argued that the Shapley
supercluster could be responsible for the anomalies 
on large angular scales (Cooray \& Seto 2005; Raki\'c et al. 2006). 
However, as shown in figure 
\ref{fig:f3}, the amplitude in the direction to the 
Shapley supercluster (in the direction of the contact point of the two hypothesized
 voids) is somewhat smaller than those of the minima in
either the quadrupole or the octopole.  
Furthermore, in the flat-$\Lambda$ FRW
model, the presence of the 
linear ISW effect (blueshift effect for mass 
concentration) can reduce the amplitude of the temperature fluctuations
due to the Rees-Sciama effect (redshift effect for mass concentration)
in the quasi-linear regime (locally $\Omega_m \sim 1$). 
Because the highly non-linear structure ($\delta \gg 1$) is 
concentrated in a particular region, 
it cannot strongly affect the large angular
fluctuations. Thus, we conclude that
such a scenario has a potential difficulty in explaining the 
anomalies in the FRW model with a cosmological constant. 
\begin{figure*}
\epsscale{0.7}
\plotone{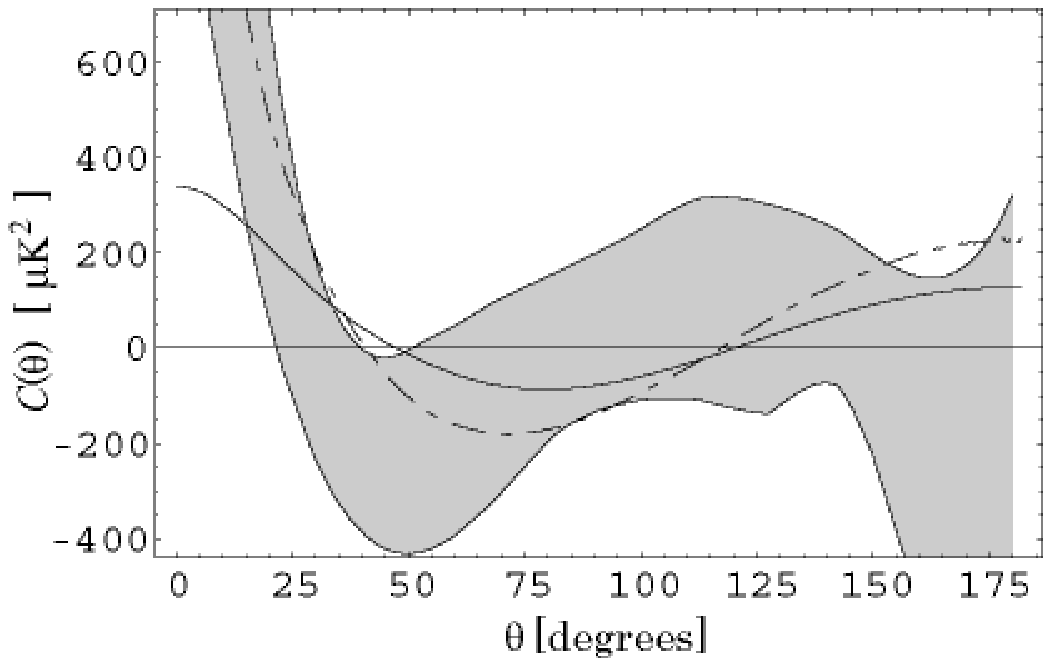}
\caption{Two-point correlation function 
$C(\theta)$ for the void model as in figure 3 (solid curve)
and for the WMAP three year data (gray region) expected for
a statistically isotropic sky with best-fit $\Lambda$CDM 
cosmology and for the best-fit flat $\Lambda$CDM model with 
$\Omega_{b,0}=0.04$ and $\Omega_{m,0}=0.24$ (dash-dotted
 curve). The gray region
denotes the maximum likelihood error at the 68\% confidence level, 
which consists of the cosmic variance and the measurement errors. 
 } \label{fig:f4}
\end{figure*}
\section{Mean contribution from dust-filled voids}
\begin{figure*}
\epsscale{1.0}
\plotone{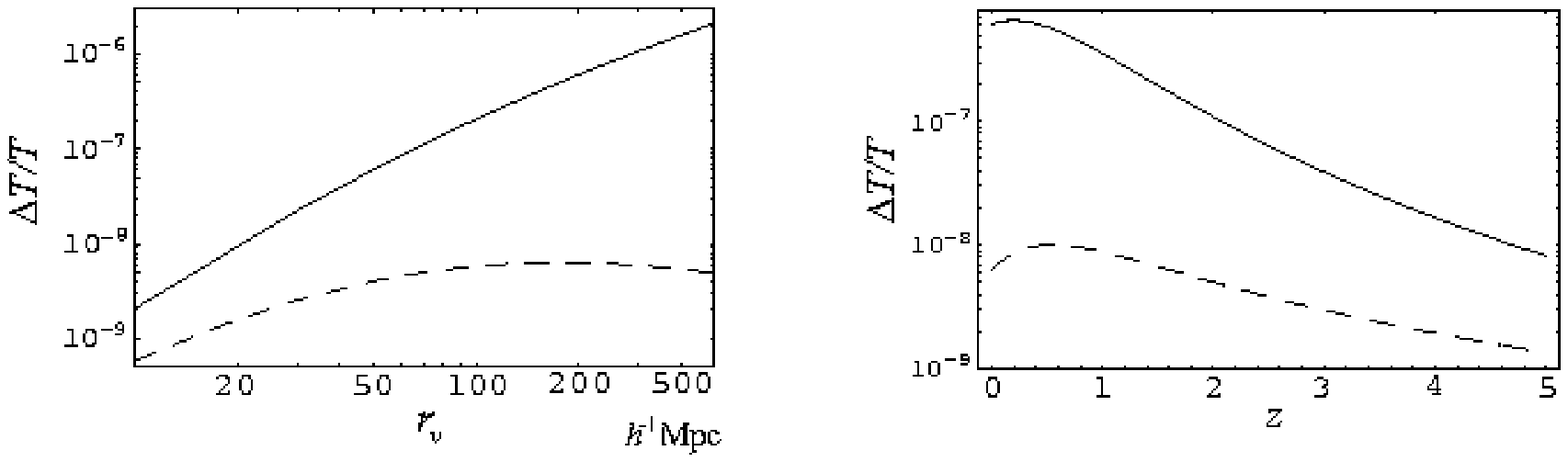}
\caption{Mean amplitudes of temperature anisotropy for a linear void
$|\delta_m| \ll 0.1$ as a function of void radius $r_v$ at $z=0$ ({\it
 {left}}),and  as a function 
of void redshift $z$ for $r_v=200 h^{-1}$ Mpc ({\it {right}}). Contribution
from the linear ISW effect is denoted by solid curves and 
that from the second-order effect is denoted by dashed curves.  
We assume $(\epsilon,\xi)=(0.0)$ and cosmological parameters
$(\Omega_{m,0},\Omega_{b,0},h,\sigma_8,n)=
(0.24,0.04,0.73,0.74,1)$ (Spergel et al. 2007).}
\label{fig:f5}
\end{figure*}

As shown in section 3, 
for linear voids $|\delta_m |\ll 1$ at $z<1$, 
the temperature anisotropy due to a dust-filled 
void in the flat-$\Lambda$ FRW background is 
proportional to $\delta_m r_v^3$ due to the 
linear ISW effect. Assuming the cold dark matter (CDM) 
power spectrum, the mean amplitude of the temperature 
anisotropy due to a linear void is proportional
to $r_v$ because $|\delta_m|_{r.m.s} \propto r_v^{-2}$ for the linear
regime. Therefore, we expect larger contributions from larger 
voids provided that $|\delta_m|_{r.m.s} \ll 1$ (solid curve 
in the right panel in figure \ref{fig:f4}). 
For quasi-linear voids $|\delta_m|_{r.m.s}=O(10^{-1})$, or linear voids
at high $z$, the second-order effect 
can also contribute to the anisotropy.
Because we have $|\delta_m|_{r.m.s} 
\propto r_v^{-1/3}$ for the non-linear
regime, the contribution from the second-order effect is 
maximum at the scale $r_v \sim 2\times 10^2 h^{-1}$ Mpc (dashed curve 
in the right panel in figure \ref{fig:f5}).
In the high redshift region $z>1$, the contribution from 
the linear ISW effect is significantly reduced because
the gravitational potential decay does not occur in the 
matter-dominant epoch. Similarly, the contribution from 
the second-order effect is also significantly reduced at $z>1$ 
(\ref{fig:f5}, {\ti{right}}).  

From these features, we conclude that 
local dust-filled quasi-linear 
voids with a radius $r_v \sim 200 h^{-1}$ Mpc  
are natural candidates for explaining 
the origin of the claimed anomalies
on large angular scales.  

However, one may argue that the possibility
of having such large voids is extremely small. 
Indeed, if we assume a flat-$\Lambda$ CDM 
cosmology, we find that a $\sim 30 \%$
mass deficiency at the scale of $r_v=200 h^{-1}$ Mpc
is only a $13 \sigma$ result for linear fluctuations with 
$\sigma_8=0.74$
or $11 \sigma$ for the preWMAP3 
normalization $\sigma_8=0.9$. 
\footnote{We consider a mean 
amplitude of linear fluctuations smoothed by a top-hat
window function with a radius $r_v=200 h^{-1}$ Mpc for
cosmological parameters $(\Omega_{m,0},\Omega_{b,0},h,\sigma_8,n)=
(0.24,0.04,0.73,0.74,1)$}
Assuming the CDM power spectrum (Bardeen et al. 1986), 
the expected temperature
anisotropy for a linear (not asymptotic) 
void with radius $r_v=200 h^{-1}$ Mpc 
at $z\sim 0$ is $\Delta T/T=O(10^{-7})$. 

This inconsistency may be circumvented by considering a 
percolation process of small non-linear voids (Colombi et al. 2000;
Hanami 2001). As $N-$body simulations
 (Colberg et al. 2005) and local observations (Patiri et al. 2006) 
suggest, at low redshifts $z<1$, 
non-linear voids with a typical radius of the 
order of  $O(10) h^{-1}$ Mpc cover almost the entire space.  
Therefore, small non-linear voids can percolate each other to make 
larger dust-filled local quasi-linear voids. 
Interestingly, a low-density region with a $25\%$ deficiency 
at $z<0.1$ over a volume of $300 h^{-1} \textrm{Mpc}^3$ 
has been discovered in the galaxy distribution in the 
south Galactic cap (Frith et al. 2003, 2005; Busswell et al. 2004). 
It seems that the underdense region
consists of two local voids in the ranges $0.03<z<0.06$ 
and $0.07<z<0.1$, suggesting that the region consists of 
several dust-filled voids with a radius $r_v \sim 50 h^{-1}$ Mpc. 
If such a low-density region can be regarded as 
a homogeneous dust-filled asymptotically evolving void,
we expect an anisotropy in the local Hubble constant 
$\delta_H \sim -\Omega_m^{0.6} \delta_m/6\sim 0.02$.
It can be even larger than 
$\delta_H =(1+f^{-1}(w))\Omega_m \delta_m/2\sim 0.04$
if it is not asymptotically evolved.
Although the substructure inside the void can also affect the
anisotropy in the Hubble constant, we expect that 
the mean value over the low-density region will show a systematic 
deviation in comparison with the global value by several
percent. Detection of a correlation between  local 
large-scale structure and the anisotropy in the local Hubble constant 
would be strong evidence for the postulated local dust-filled large voids. 

\section{Summary and discussion}
In this paper, we have explored the large angular scale temperature 
anisotropy owing to the homogeneous local dust-filled 
voids in the flat-$\Lambda$ FRW universe. 
A compensated asymptotically evolving 
dust-filled void with a comoving radius 
$r_v=(2-3) \times 10^2h^{-1}$ Mpc and a 
density contrast $\delta_m=-0.3$ will give a 
cold spot with anisotropy $\Delta T/T \sim -10^{-5}$.
A pair of such compensated local voids  
in the direction $(l,b)=(-30^\circ,-30^\circ)$ might
explain the quadrupole/octopole alignment and the planarity of
quadrupole and octopole. 
As in the void model in the EdS background,
the temperature anisotropy due to the dust-filled homogeneous void
is proportional to $(r_v/H^{-1})^3$. 
For asymptotically evolving voids, contributions 
from the term that is proportional to $\delta_m^2$ 
always enhance the linear ISW effect which is proportional to $\delta_m$. 
However, for local voids that expand sufficiently faster than the 
asymptotic velocity of the wall, 
contributions from the term $\propto\delta_m^2$ 
can suppress the linear ISW effect. 
Thus the enhancement/suppression depends on the non-linear dynamics
of the void. We have also shown that the contribution to the CMB 
anisotropy from the linear ISW and the second order effects is 
more significant for local voids at $z<1$ than those
at high redshifts $z>1$.

Assuming the CDM power spectrum based on the standard
inflationary scenario, the expected amplitude 
of the temperature anisotropy owing to the linear void
with radius $r_v=2\times 10^2h^{-1}$ Mpc 
is just $\Delta T/T=O(10^{-7})$. However, at low redshifts $z<1$, 
percolation of small non-linear voids could
make large quasi-linear voids with a 
density contrast $|\delta_m|\sim 0.1$.
The observed low-density region with a $25\%$ deficiency 
at $z<0.1$ over a volume of $300 h^{-1} \textrm{Mpc}^3$ 
in the south Galactic cap (Busswell et al. 2004) 
might be evidence of 
such a structure. 
Detailed observational 
properties can be obtained by ongoing projects such as the 
6dF galaxy survey (Heath et al. 2004).
On the theoretical side, more elaborate modeling of the void percolation
process is required because the signature on the CMB may depend on the configuration
and the momentum of the substructure inside the void. Determination 
of non-linear parameters $(\epsilon, \eta)$ based on more realistic 
models is also an important task. 
Once these observational and theoretical issues are settled,
we can determine whether new physics, such as 
primordial non-Gaussianity (Mathis et al. 2004) 
or an enhancement in the gravity in the dark energy sector
 (Farrar \& Rosen 2006), may be required.

Another implication unique to the dust-filled large voids 
postulated here concern increased dispersion in the locally measured 
Hubble constant as measured both in different directions and at different
redshifts. As we have shown, for voids with a density contrast 
$\delta_m=-0.3$, the expected fluctuation in the Hubble
constant is as large as $2-4 \%$ (Tomita 2000, 2001). 
Although, in this paper, we have assumed that we are outside the 
dust-filled voids, it is also possible that 
we are inside the void.  If the distance between the Milky Way
and the wall of the void is sufficiently small, 
the ratio of our peculiar 
velocity to the light velocity will be $v/c=O(10^{-3})$ 
for a radius $r_v=(2-3) \times 10^2h^{-1}$ Mpc,
and a density contrast $\delta_m=-0.3$. Therefore, the 
order of the predicted amplitude of the dipole is similar to that  
of the observed CMB dipole (Tomita 2003). In this case,
we expect a positively skewed dispersion in the Hubble constant
$\delta_H\sim 0.02-0.04$ in the direction to the center of the void
 (approximately at $(l,b)=(-30^\circ, -30^\circ)$). 
Accuracy measurement of $\sim 1 \%$ in the Hubble constant would be  
achieved by increasing the number of samples of Type Ia supernovae 
from $N\sim 50 $ to $5000$ provided that the error decreases as
$1/\sqrt{N}$. 

Another approach to peculiar velocities uses galaxy surveys that 
apply different  distance calibrators 
(Pike \& Hudson 2005; Watkins and Feldman 2007).
The most optimistic individual galaxy distance estimates cited, 
using $I$-band surface brightness 
fluctuations,  have an accuracy of 10\% for half the sampled galaxies
and extend to 40$h^{-1}$Mpc (Tonry et al. 2001). 
Other studies go deeper, for example the Tully-Fisher relation has been 
applied to a sample of 3000 spiral galaxies out to  200$h^{-1}$Mpc 
(Karachentsev et al. 2006) with distance errors of 15\%.
It should be possible to apply these techniques to test our hypothesis 
by targeting several thousand galaxies that sample the peculiar
velocities induced by our nearest postulated voids and their associated
shells, which extend over thousands of square degrees.

Thus further study of correlation between the 
large scale structure and
the variation in the locally measured Hubble 
constant is necessary in order to prove/disprove our 
void scenario. 

Our findings may shed new light on the issue of
cross-correlation between the large-scale structure and the CMB.
Within a flat $\Lambda$CDM model, recent works on the cross-correlation 
suggest a best-fit value $\Omega_\Lambda \sim 0.85$ (Rassat et
al. 2006; Cabre et al. 2006), which 
is larger than the best-fit value $\Omega_\Lambda=0.70$
from the joint analysis of the CMB and 
the large-scale structure (Tegmark et al. 2004).
Because the second order effect 
in quasi-linear asymptotic voids systematically 
enhances the linear ISW effect,
the observed deviation could be explained by such
quasi-linear or non-linear local voids in the flat-$\Lambda$ universe  
with $\Omega_\Lambda=0.70$.

If the low$-\ell$ CMB anomalies persist, our phenomenological model
provides a simple astrophysical interpretation with no recourse to new
physics. Perhaps the most promising tests of our hypothesis will come
from improved sampling of Hubble flow anisotropies. If these
directions coincide with the locations of our postulated voids, the
case becomes more compelling. CMB polarisation maps and lensing
studies could potentially reveal interesting signals. For example, the
inflow pattern in the void wall will induce a small polarisation
signal, as will the associated gravitational lensing of the CMB. 
The expected amplitude of the scattering angle of photons due to the postulated
large local voids is $\sim \delta_m^2 \xi^2 \sim 0.1^\circ $,  
corresponding to subdegree scales. These effects are
small, amounting to an imprint on the ambient polarisation 
pattern of order a percent ($\sim v/c$), but the phase
structure would be unique and correlated with both the temperature map
and the large-scale galaxy distribution.  The global mean value of the
Hubble constant will be slightly reduced, as the required dominance of
voids biases the measured local $H_0$ to be slightly high. Although
the effect is small, it has potential significance for future dark
energy surveys, which will combine CMB studies with deep galaxy
redshift surveys. For example, the degeneracy between the 
Hubble constant and the determination of 
the curvature of the universe using the CMB 
will need to be taken into account.

One anomaly we have not sought to incorporate explicitly is the 
north-south power excess
in the CMB.  It may be tempting to attribute this to a directional 
deficiency of local power,
as has been previously noted in deep galaxy counts (Busswell et
al. 2004) and which in the present context consists of a  compensated
void complex that correlates the CMB and 
large-scale structure signals.

The reduction in normalisation
of $\sigma_8$ to 0.74 makes it more difficult to account for large-scale 
features in the galaxy distribution without introducing extra
large-scale power. Our example of tuned voids is one manifestation of
extra power that would inevitably, if incorporated into the
cosmological initial conditions of a simulation, boost the
significance of large-scale features such as the SDSS ``Great Wall''
(Gott et al. 2005) and rare massive clusters (Baugh et al. 2004).

\acknowledgements

We thank K. Tomita and N. Sakai for useful discussions and 
comments. This work is in part supported by a Grant-in-Aid for
Young Scientists (17740159) from the Ministry of Education, Culture,
Sports, Science and Technology in Japan.
\vspace{5cm}
\appendix
\section{Linear perturbation of Hubble parameter}
In what follows, we calculate the Hubble parameter contrast at the 
matter- or $\Lambda$-dominant era in the linear order.  
Let us write the Friedmann equation 
at a point $\x$ in a perturbed universe in terms of
the scale factor $a$, the Hubble parameter $H$, the matter density 
$\rho_m$, the gravitational constant $G$, 
the curvature $K$, and the cosmological
constant $\Lambda$ as 
\BE
H^2(\x,t)=\f{8 \pi G}{3}\rho_m(\x,t)-\f{K(\x,t)}{a^2(t)}+\f{\Lambda}{3}
\label{A1}
\EE
where the Hubble parameter $H$ is defined on a comoving slice and
the scale factor at the present time $t_0$ is assumed to be $a_0=1$.
In what follows, we introduce the total density
$\rho= \rho_m+\f{\Lambda}{8 \pi G}$.
The linear perturbation component in equation (\ref{A1}) in 
the Fourier space for the flat FRW background can be written as
\BE
2 \bar{H} \delta H_\K=\f{8 \pi G}{3}\delta 
\rho_\K-\f{2}{3}\biggl(\f{k}{a}  \biggr)^2 {\cal{R}}_\K, 
\label{A2}
\EE
where $\bar{H}$ is the spatially averaged 
Hubble parameter and the curvature perturbation ${\cal{R}}$ is 
defined as  
\BE
\delta K_\K=\f{2}{3} k^2 {\cal{R}}_\K. \label{A3}
\EE 
If we can neglect the anisotropic stress, the relation between the 
Newtonian gravitational potential $\Phi_\K$ and 
the total energy density contrast $\delta_\K $ is given by
the Poisson equation
\BE
- \biggl(\f{k}{a} \biggr)^2   \Phi_\K=4 \pi G \delta \rho_\K. \label{A4}
\EE
The time evolution of $\delta$ is given by
the continuity equation
\BE
\dot{\delta \rho}_\K=-3(\bar{\rho}+\bar{P})\delta H_\K-3 \bar{H}
\delta \rho_\K, \label{A5}
\EE
where $\bar{\rho}$ and $=\bar{P}$ are the spatially averaged
total energy density and total pressure, respectively.
The perturbed Raychaudhuri equation is (Liddle \& Lyth 2000)
\BE
\dot{\delta H}_\K=-2 \bar{H} \delta H_\K-\f{4 \pi G}{3}\delta \rho_\K
+\f{1}{3}\biggl(\f{k}{a} \biggr)^2 \f{\delta P_\K}{\bar{\rho}+\bar{P}}.
\label{A6}
\EE
From equations (\ref{A2}-\ref{A6}), we have
\BE
\f{2}{3}\bar{H}^{-1}\dot{\Phi}_\K+\f{5+3w}{3}
\Phi_\K=-(1+w){\cal{R}_\K},
\label{A7}
\EE
where the effective equation of state $w=\bar{P}/\bar{\rho}$
is written in terms of the density parameter for the matter 
$\Omega_{m,0}$ and that for the cosmological constant 
$\Omega_{\Lambda,0}$ 
at the present time $t_0$ as
\BE
w=-\f{1}{\Omega_{m,0}/(\Omega_{\Lambda,0}~ a^3)+1}. \label{A8}
\EE
From the time derivative of the 
local Friedmann equation (\ref{A2}) multiplied by $a^2$, 
the continuity equation (\ref{A5}), and 
the Raychaudhuri equation (\ref{A6}),
we have ${\dot{\cal{R}}_\K}=- \bar{H} \delta P_\K/(\bar{\rho}+\bar{P})=0$ 
because the pressure perturbation vanishes $\delta P_\K=0$ at the 
matter- or $\Lambda$- dominant era.  
Plugging $\Phi_\K=f(w){\cal{R}_\K}$ into equation (\ref{A7}),  
since ${\dot{\cal {R}}_\K}=0 $, $f(w)$ satisfies  
\BE
2w(1+w)\f{df}{dw}+\f{5+3w}{3}f+1+w=0. \label{A9}
\EE
The solution is 
\BE
f(w)=-\f{3}{5}(1+w)^{1/3}{}_2 F_1\biggl[ \f{5}{6},\f{1}{3},\f{11}{6},-w
\biggr ], \label{A10}
\EE
which satisfies the boundary condition 
$f=-3/5$ for $w=0$. Plugging $\Phi_\K=f(w){\cal{R}_\K}$ into
equation (\ref{A2}) and (\ref{A4}), we have
\BE
\f{\delta H_\K}{\bar{H}}=\f{1+f^{-1}(w)}{2}\delta_\K, \label{A11}
\EE
which yields
\BE
\delta_H=\f{1+f^{-1}(w)}{2}\delta_m \Omega_m, \label{A12}
\EE
where the evolution of $f(w)$ is given by equation(\ref{A8}) and (\ref{A10}).
The equation (\ref{A12}) is applicable to the matter- or $\Lambda$-
dominated era at which the anisotropic stress is negligible.
\section{Linear integrated Sachs-Wolfe effect}
We evaluate the magnitude of the linear integrated 
Sachs-Wolfe (ISW) effect, which comes from the time evolution
of the gravitational potential as the photon traverses
the void. 
 
In terms of the gravitational potential $\Psi$ , 
contribution from the ISW effect is approximately evaluated as 
\BE
\f{\Delta T}{T}\sim  \f{\del \Psi}{\del t} \Delta t, \label{ISW1}
\EE
where $\Delta t$ is the duration time for which 
the photons pass through the void. From the Poisson equation, in the
Newtonian limit, $\Delta t \ll H_0^{-1}$,  the gravitational potential
can be written in terms of the density contrast 
$\delta_\K=\delta \rho_\K/\rho$ 
inside the void in the Fourier space as 
\BEA
\dot{\Psi}_\K
&\sim & -\f{1}{k^2} \f{\del}{\del t}(H^2 a^2 \delta_\K)
\nonumber
\\
&=&
-\f{1}{k^2} (2 \dot{a} \ddot{a} \delta_\K+\dot{a}^2 \dot{\delta}_\K)
\label{eq:B2}
\EEA
where  a dot means the partial time derivative 
$\cdot \equiv \f{\del}{\del t}$ and $k$ is the wavenumber.  
Plugging the peculiar velocity inside the void 
at comoving distance $r$ from the center $v=\delta_H H a r$
into the continuity equation
\BE
\f{\del \delta}{\del t}+a^{-1} \nabla \cdot \bold{v}=0,
\EE
we have 
\BE
\dot{a}^2 \delta_\K^2=-3 H^3 a^2 \alpha \delta_\K,
\label{eq:B4}
\EE
where $\alpha=(1+f^{-1}(w))/2$. Writing 
the wavenumber in terms of the void comoving 
radius $r_v$ as $k \sim \pi/r_v$, equation (\ref{eq:B2}) 
and equation (\ref{eq:B4}) yield,
\BE
\f{\Delta T}{T} \sim -\f{1}{\pi^2} \xi^3 (2 q+3 \alpha )\delta,
\label{B5}
\EE
where $\xi=a r_v/H^{-1}$ and $q$ is the deceleration parameter. 
Thus, the anisotropy owing to the linear ISW effect is proportional
to the density contrast $\delta$ and $\xi^3$. 
In terms of the matter density contrast
$\delta_m=\delta \rho_m/\rho_m$ and the
matter density parameter $\Omega_m$, 
equation (\ref{B5}) can be written as 
\BE
\f{\Delta T}{T}\sim s\xi^3 \delta_m\Omega_m, 
\label{B6}
\EE
where 
\BE
s=-\f{1}{\pi^2} \biggl( 3 \Omega_m -\f{1}{2}-
\f{5}{2}\Omega_m^{-1/3}F^{-1} \biggr).
\EE

\end{document}